\documentclass[12pt,letter]{article}
\pdfoutput=1
\usepackage{graphicx, epsfig, color,cite}
\usepackage{amsmath}
\usepackage{amssymb}
\usepackage{float}
\usepackage{caption,subcaption,graphicx}
\usepackage{hyperref}

\def\eslt{\not\!\!\!{E_T}}
\def\to{\rightarrow}

\def\bi{\begin{itemize}}
\def\ei{\end{itemize}}

\def\tchi{\tilde\chi}

\def\tb{\tilde b}

\def\tst{\tilde t}

\def\tg{\tilde g}

\def\tq{\tilde q}
\def\alt{\lesssim}
\def\agt{\gtrsim}
\def\be{\begin{equation}}  
\def\ee{\end{equation}}  
\def\bea{\begin{eqnarray}}  
\def\eea{\end{eqnarray}}

\begin{document}
\begin{titlepage}
\begin{flushright}
OU-HEP-230701
\end{flushright}

\vspace{0.5cm}
\begin{center}
  {\Large \bf Top squarks from the landscape at high luminosity LHC
    }\\
\vspace{1.2cm} \renewcommand{\thefootnote}{\fnsymbol{footnote}}
{\large Howard Baer$^{1}$\footnote[1]{Email: baer@ou.edu },
Vernon Barger$^2$\footnote[2]{Email: barger@pheno.wisc.edu},
Juhi Dutta$^1$\footnote[3]{Email: juhi.dutta@ou.edu},\\
Dibyashree Sengupta$^3$\footnote[4]{Email: Dibyashree.Sengupta@lnf.infn.it} and
Kairui Zhang$^2$\footnote[5]{Email: kzhang89@wisc.edu}
}\\ 
\vspace{1.2cm} \renewcommand{\thefootnote}{\arabic{footnote}}
{\it 
$^1$Homer L. Dodge Department of Physics and Astronomy,
University of Oklahoma, Norman, OK 73019, USA \\[3pt]
}
{\it 
$^2$Department of Physics,
University of Wisconsin, Madison, WI 53706 USA \\[3pt]
}
{\it
  $^3$ INFN, Laboratori Nazionali di Frascati,
Via E. Fermi 54, 00044 Frascati (RM), Italy}

\end{center}

\vspace{0.5cm}
\begin{abstract}
\noindent
Supersymmetric models with low electroweak finetuning are expected to be more
prevalent on the string landscape than finetuned models.
We assume a fertile patch of landscape vacua containing the
minimal supersymmetric standard model (MSSM) as low energy/weak scale
effective field theory (LE-EFT).
Then, a statistical pull by the landscape to large soft terms is balanced by
the requirement of a derived value of the weak scale which is not too far from
its measured value in our universe.
Such models are characterized by light higgsinos in the few hundred GeV
range whilst top squarks are in the 1-2.5 TeV range with large trilinear 
soft terms which helps to push $m_h\sim 125$ GeV.
Other sparticles are generally beyond current LHC reach and the
$BR(b\to s\gamma )$ branching fraction is nearly equal to its SM value.
The light top-squarks decay comparably via
$\tst_1\to b\tchi_1^+$ and $\tst_1\to t\tchi_{1,2}^0$ yielding mixed
final states of $b\bar{b}+\eslt$, $t\bar{b}/\ \bar{t}b +\eslt$ and
$t\bar{t}+\eslt$.
We evaluate prospects for top squark discovery at high-luminosity (HL) LHC
for the well-motivated case of natural SUSY from the landscape.
We find for HL-LHC a $5\sigma$ reach out to $m_{\tst_1}\sim 1.7$ TeV
and a 95\% CL exclusion reach to $m_{\tst_1}\sim 2$ TeV.
These reaches cover {\it most} (but not all) of the allowed stringy
natural parameter space!
\end{abstract}
\end{titlepage}

\section{Introduction}
\label{sec:intro}

The lightest supersymmetric (SUSY) partner of the top quark,
the so-called top-squark $\tst_1$, has for long been a lucrative target
for supersymmetry searches at hadron colliders.
Early estimates of EENZ/BG\cite{Ellis:1986yg,Barbieri:1987fn} naturalness,
using the measure $\Delta_{BG}\equiv max_i|\frac{p_i}{m_Z^2}\frac{\partial m_Z^2}{\partial p_i}|<\Delta_{BG}(max)$ (where the $p_i$ are taken as fundamental
theory parameters, usually assumed to be a set of high scale soft
SUSY breaking terms) found $m_{\tst_1}\alt 300-400$ GeV for
$\Delta_{BG}<10-30$\cite{Dimopoulos:1995mi}.
Using an alternative measure $\delta m_h^2/m_h^2 <\Delta_{HS}$,
it was expected that {\it three} third generation squarks should all have
mass $m_{\tst_{1,2},\tb_1}\alt 500$ GeV\cite{Kitano:2006gv,Papucci:2011wy,Brust:2011tb}.
These theoretical naturalness computations may be compared to recent limits
from LHC searches where both ATLAS\cite{ATLAS:2020dsf,ATLAS:2020xzu}
and CMS\cite{CMS:2021beq} find that
$m_{\tst_1}\agt 1.2$ TeV from $pp$ collisions at $\sqrt{s}=13$ TeV and with
$\sim 139$ fb$^{-1}$ of integrated luminosity.
Taken at face value, this confrontation between theory and experiment would
indicate that the paradigm of weak scale supersymmetry\cite{Baer:2006rs}
is highly implausible as a route to physics beyond the Standard Model
(SM)\cite{Dine:2015xga}.

One resolution to the supersymmetry naturalness conflict is that the early
theoretical naturalness calculations turned out to be large
overestimates of the actual finetuning\cite{Baer:2013gva,Baer:2023cvi}.
For the BG measure, it is emphasized in
Ref. \cite{Baer:2013gva,Mustafayev:2014lqa,Baer:2014ica,Baer:2023cvi}
that the fundamental theory parameters $p_i$ should not be taken as a set
of independent soft SUSY breaking terms, since in any more UV-complete theory,
these are all correlated. For example, in gravity-mediation SUSY breaking models with a well-specified SUSY breaking sector,
then the soft terms are all computed as multiples of the gravitino mass
$m_{3/2}$.\footnote{For instance, in dilaton-dominated SUSY breaking, then $m_0=m_{3/2}$
  with $A_0=-m_{1/2}=\sqrt{3}m_{3/2}$\cite{Brignole:1993dj}.}
Adopting independent soft terms as the $p_i$ just parametrizes our ignorance
of the SUSY breaking mechanism, but can lead to overestimates of finetuning by
up to three orders of magnitude\cite{Baer:2023cvi}.
Alternatively, the $\Delta_{HS}$ measure attempts to tune dependent quantities
$m_{H_u}^2$ and $\delta m_{H_u}^2$ one against the other, which again leads to
up to three orders of magnitude overestimates of finetuning\cite{Baer:2023cvi}.
If one instead adopts the more conservative {\it electroweak} finetuning measure
$\Delta_{EW}$\cite{Baer:2012up,Baer:2012cf},
then top-squark masses are allowed up to several TeV at little cost to
finetuning since their contributions to the weak scale are suppressed by
loop factors (for a recent review, see {\it e.g.} Ref. \cite{Baer:2020kwz}).

In most supersymmetric models of particle physics--
even in the case of high scale scalar mass universality--
the lighter top squark $\tst_1$ is expected to be the lightest of all the
squarks.
It thus presents a lucrative target for 
supersymmetry discovery at hadron collider experiments such as the CERN LHC.
The lightness of the top squark, relative to other squarks, arises from two reasons: 
1. the large top-quark Yukawa coupling $f_t$ acts to drive top squark soft terms to lower 
values than other squarks (assuming an initial degeneracy amongst all squark soft terms at the high scale) and 
2. the large top Yukawa enhances the mixings amongst the top squarks, and
large mixing typically acts to further split the top squark eigenmasses,
driving the lighter one down and the heavier stop
$\tst_2$ to larger values (relative to the no mixing case). 

A third effect arises from the string landscape picture\cite{Bousso:2004fc,Susskind:2003kw,Douglas:2006es}.
In the string landscape, where of order $10^{500}$ vacua solutions\cite{Ashok:2003gk} arise from compactification
from 10 to 4 spacetime dimensions, then each vacuum solution corresponds to a different set
of $4-d$ low energy effective field theory law of physics.
The string landscape provides a natural setting for Weinberg's anthropic
solution to the cosmological constant problem\cite{Weinberg:1987dv}
in an eternally inflating multiverse. 
If similar reasoning is applied to the origin of the SUSY breaking scale, then it is expected 
that no particular value of the (complex-valued) SUSY breaking $F$ terms or
(real-valued) SUSY breaking $D$-terms are favored over any other.
In that case, then on rather general grounds, the landscape is expected to 
statistically favor large soft terms via a power law\cite{Douglas:2004qg}
\be
f_{SUSY}\sim m_{soft}^{2n_F+n_D-1}
\ee 
where $f_{SUSY}$ encodes the expected statistical distribution of landscape soft terms. 
Thus, even the textbook case of SUSY breaking via a single $F$-term field
would yield a linear draw to large soft terms. 

Naively, one might expect such a distribution to favor high scale SUSY breaking. 
However, the weak scale soft terms and SUSY-preserving $\mu$ parameter
determine the magnitude of the weak scale via the scalar potential
minimization conditions under the radiative breaking of
electroweak symmetry:
\be
m_Z^{PU2}=\frac{m_{H_d}^2+\Sigma_d^d-(m_{H_u}^2+\Sigma_u^u)\tan^2\beta}{\tan^2\beta -1}-\mu_{PU}^2
\label{eq:mzsPU}
\ee 
where the label $PU$ stands for parameter values in each separate {\it pocket universe} 
within the greater multiverse.
Here, following Weinberg, we assume a so-called 
fertile patch of the multiverse wherein the low energy/weak scale
effective field theory (LE-EFT) consists of the MSSM 
(plus some additional fields such as a PQ sector) but with variable soft terms
and hence variable values for the associated weak scale $m_{weak}^{PU}$. 
The value of $m_{weak}^{PU}$ is typically $\ne m_{weak}^{OU}$, where $OU$ stands for 
a quantity's value in {\it our universe}.
Agrawal {\it et al.} (ABDS)\cite{Agrawal:1997gf} have shown that for complex
nuclei-- and hence atoms as we know them-- to form in a PU, that the value
of $m_{weak}^{PU}$ must lie within the ABDS window, typically
$m_{weak}^{PU}\sim  (0.5- 5) m_{weak}^{OU}$
(the atomic principle).
The ABDS anthropic window thus vetoes vacua with improper 
electroweak symmetry breaking (EWSB),
such as solutions with no EWSB or charge or color breaking (CCB) minima; it also excludes the vast majority of high scale SUSY solutions which typically lead to $m_{weak}^{PU}$ far beyond the ABDS window.
The string landscape approach to soft SUSY breaking within the MSSM has led
to some success in that it statistically predicts a Higgs boson mass
$m_h\simeq 125$ GeV whilst sparticles
are typically well beyond current LHC search bounds\cite{Baer:2017uvn}.

Returning to top squarks, the large value of the top quark Yukawa coupling
enhances the radiative correction terms $\Sigma_u^u(\tst_{1,2})$ in
Eq. \ref{eq:mzsPU} relative to $\Sigma_u^u(\tq_i)$
(where $\tq_i$ denotes squark masses of the first two generations. 
Thus, for independent soft terms for each generation (as is generic in gravity
mediation\cite{Soni:1983rm,Kaplunovsky:1993rd,Brignole:1993dj}) 
then the squark and slepton masses of the first two generations will get
pulled to much higher values, typically $m_{\tq_i}\sim 10-40$ TeV
(providing a mixed decoupling/quasi-degeneracy
solution to the SUSY flavor and CP problems\cite{Baer:2019zfl})
whilst $m_{\tst_1}\sim 1-2.5$ TeV.
As such, the string landscape provides additional strong motivation for
top-squark pair searches at LHC as compared to other sparticle searches
(although the search for light higgsinos with $m(higgsino)\sim 100-400$ GeV
is also especially lucrative\cite{Baer:2011ec,Han:2014kaa,Baer:2014kya,Han:2015lma,Baer:2020sgm,Baer:2021srt}).

In this paper, after a brief review of some previous relevant works in Subsec. \ref{ssec:review}, 
in Sec. \ref{sec:stops} we will present landscape predictions for some of the 
relevant properties of light top squarks as derived from string landscape
predictions with a simple $n=1$ power law draw to large soft terms. 
We will find that while large stop mixing terms $m_tA_t$ are expected at the
weak scale (and indeed these help boost up the light Higgs mass to
$m_h\sim 125$ GeV), the lighter top squark $\tst_1$ is still typically mainly
a right-top-squark (assuming high scale degeneracy of left and right top squark
soft terms $m_{\tst_L}$ and 
$m_{\tst_R}$, as expected by intragenerational degeneracy since the elements of each generation
fill out the 16-dimensional spinor-rep of $SO(10)$). Also, we will find that
the branching fraction $BR(b\to s\gamma )$ is expected to be very near its Standard Model (SM) value (in agreement with data and in accord with the general
expectation for TeV-scale top-squarks).
We will also determine the expected $\tst_1$ branching fractions
which will determine the associated LHC search signatures.
In Sec. \ref{sec:LHC}, we examine top-squark pair production rates and expected signal channels
which are expected for HL-LHC searches.
In Sec. \ref{sec:BM} we introduce a natural top-squark benchmark point and associated model line. 
In Sec. \ref{sec:reach}, we give cuts and $m_{T_2}$ distributions for each of the the three major signal channels. By combining results, we present $5\sigma$ reach and 95\%CL exclusion limits versus
$m_{\tst_1}$. 
In Sec. \ref{sec:conclude}, we present a brief summary and conclusions from our
results.

\subsection{A brief review of some previous relevant works}
\label{ssec:review}

The first few papers on top-squark phenomenology focussed
on the possibility for $t\to \tst_1\tchi_i^0$ decays which could
disrupt top-quark discovery signatures at the CERN S$p\bar{p}S$\cite{Bigi:1985aq,Baer:1985hd} and Fermilab Tevatron colliders\cite{Baer:1991cb}.
Direct top-squark pair production at the Tevatron within the framework
of simplified models was already examined in Ref. \cite{Baer:1994xr}
shortly before the actual discovery of the top-quark.
In Ref. \cite{Rolbiecki:2009hk}, the capability of LHC to measure the top-quark mixing angle $\theta_t$ was examined: the strategy promoted was to as best as one can measure the various top-squark branching fractions into different decay modes
which depend on stop mixing.
In Ref. \cite{Brummer:2012ns}, the scenario of maximal stop mixing, which
provides an explanation for the rather high Higgs mass $m_h\simeq 125$ GeV,
was examined with a view towards resolving the apparent tension between
naturalness and the light Higgs mass.
In Ref. \cite{Graesser:2012qy}, Graesser and Shelton examined top squark pair
production followed by mixed top-squark decay modes $\tst_1\to b\tchi_1^+$ with
$\tst_1^*\to \bar{t}\tchi_1^0$ and suggested a new search variable
$t=$topness to aid in identifying top jets in the final state.
In Ref. \cite{Baer:2016bwh}, it is argued that conventional finetuning measures
overestimated the severeness of top-squark mass upper bounds and instead
examined implications of the $\Delta_{EW}$ measure for top-squark properties.
Using $\Delta_{EW}\alt 30$, then top squarks may range up to $m_{\tst_1}\alt 3$
TeV provided there is a rather large weak scale $A_t$ soft term mixing
value which also elevates the Higgs mass $m_h\to\sim 125$ GeV. Thus, there
exists a significant portion of natural SUSY parameter space that lies beyond
ATLAS/CMS limits as displayed in the $m_{\tst_1}$ vs. $m_{\tchi_1^0}$
simplified model parameter plane.
In C. Han {\it et al.}\cite{Han:2016xet},
the authors recast various ATLAS/CMS top squark search results into the
$m_{\tst_1}$ vs. $m_{\tchi_1^0}$ top-squark search plane and compare against
naturalness using $\Delta_{EW}$.
In Ref. \cite{Bai:2016zou}, assuming $\tst_1\to t\tchi_1^0$ decay,
Bai {\it et al.} impose a very strong $\eslt$ cut against which two top jets
merge.
Cuts on the resulting configuration boost signal over background by 40\%
over conventional analyses.

In Ref. \cite{ATLAS:2018zrp}, the ATLAS Collaboration examined the reach of
HL-LHC for top-squark pair production followed by $\tst_1\to t\tchi_1^0$
in the top-squark search plane: for light $m_{\tchi_1^0}$,
they find using LHC14 with 3000 fb$^{-1}$ a $5\sigma$ reach to
$m_{\tst_1}\sim 1.25$ TeV and a 95\% CL exclusion to $m_{\tst_1}\sim 1.7$ TeV.
Similar results from CMS are shown in Ref. \cite{CidVidal:2018eel}.

\section{Properties of top squarks from the landscape}
\label{sec:stops}

\subsection{Scan over landscape}

In this Section, we wish to explore the predictions from the string landscape
for top squark properties. To this end, we will generate the
distribution
\be
dN_{vac}=f_{SUSY}\cdot f_{EWSB}\cdot f_{cc}\cdot dm_{soft}
\label{eq:dNvac}
\ee
where $dN_{vac}/dm_{soft}$ stands for the distribution of string vacua with
respect to the soft SUSY breaking parameters. We will assume a fertile patch
of the string landscape where the LE-EFT consists of the MSSM
(possibly augmented with a PQ sector which is only relevant for dark matter
considerations), but where in the string landscape the various soft
terms
\be
m_0(1,2),\ m_0(3),\ m_{1/2},\ A_0,\ \tan\beta,\ m_A,\ \mu 
\ee
will scan independently\cite{Baer:2020vad}.
(Note that while the soft terms would all be correlated and hence
{\it dependent} in our universe,
as discussed earlier, they should scan independently within the various
pocket universes within the greater multiverse.)
The various independent soft terms scan as a power-law
\be
f_{SUSY}\sim m_{soft}^{2n_F+n_D-1}
\ee
where $n_F$ is the number of hidden sector $F$-breaking fields
(distributed as complex numbers) and $n_D$ is the number of $D$-breaking fields
(distributed as real numbers).
For simplicity, we will adopt the textbook case $n_F=1$, a single $F$ breaking field, and $n_D=0$ so that $f_{SUSY}\sim m_{soft}^1$, {\it i.e.} a linear statistical draw
to large soft terms. We will take the non-soft term $\tan\beta$ to scan
uniformly and will fix the $\mu$ parameter to a natural value
$\mu =200$ GeV. The cosmological constant selection embedded in $f_{cc}$
does not impact the soft term selection as emphasized
by Douglas\cite{Douglas:2004qg}.

For the (anthropic) selection $f_{EWSB}$, we require the derived value of
the weak scale in each pocket universe to have 1. no charge or color breaking minima (no CCB), 2. an appropriate breakdown of EW symmetry to $U(1)_{EM}$ and
3. a derived value for the pocket universe lies within the ABDS window\cite{Agrawal:1997gf},
{\it i.e.} that $m_{weak}^{PU}\alt (0.5- 5)m_{weak}^{OU}$.
To be precise, in a pocket universe with no finetuning, this corresponds to
$\Delta_{EW}\alt 30$ since $m_{weak}^{PU}\sim m_Z\sqrt{\Delta_{EW}/2}$.

By combining the various effects in Eq. \ref{eq:dNvac}, we are able to obtain
a measure of what Douglas calls {\it stringy naturalness}\cite{Douglas:2012bu,Baer:2019cae}. While stringy naturalness is not measured by a number, we can measure it via a scan over SUSY model soft terms in accord with Eq. \ref{eq:dNvac}.
Here, we implement the $n\equiv 2n_F+n_D-1 =1$ linear scan over the
NUHM3\cite{Baer:2005bu} parameter space:
\begin{itemize}
\item $m_0(1,2): 0.1-45\ {\rm TeV}$,
\item $m_0(3):\ 0.1-10\ {\rm TeV}$,
\item $m_{1/2}:\ 0.5-3\ {\rm TeV}$,
\item $A_0:\ 0-(-20)\ {\rm TeV}$,
\item $\tan\beta :\ 3-60$ (uniform scan),
  \item $m_A:\ 0.3-10\ {\rm TeV}$,
  \end{itemize}
with $\mu$ fixed at a natural value of 200 GeV.\footnote{The
  SUSY conserving $\mu$ parameter arises from whatever solution to the SUSY $\mu$ problem is imposed.
  For a review of twenty solutions to the SUSY $\mu$ problem,
  see Ref. \cite{Bae:2019dgg}.}
For each set of input parameters in the NUHM3 model, we use Isajet 7.88\cite{Paige:2003mg}
to compute the corresponding sparticle and Higgs boson masses and other properties.

\subsection{Top squark mass and mixing}

\begin{figure}[htb!]
\begin{center}
  \includegraphics[height=0.3\textheight]{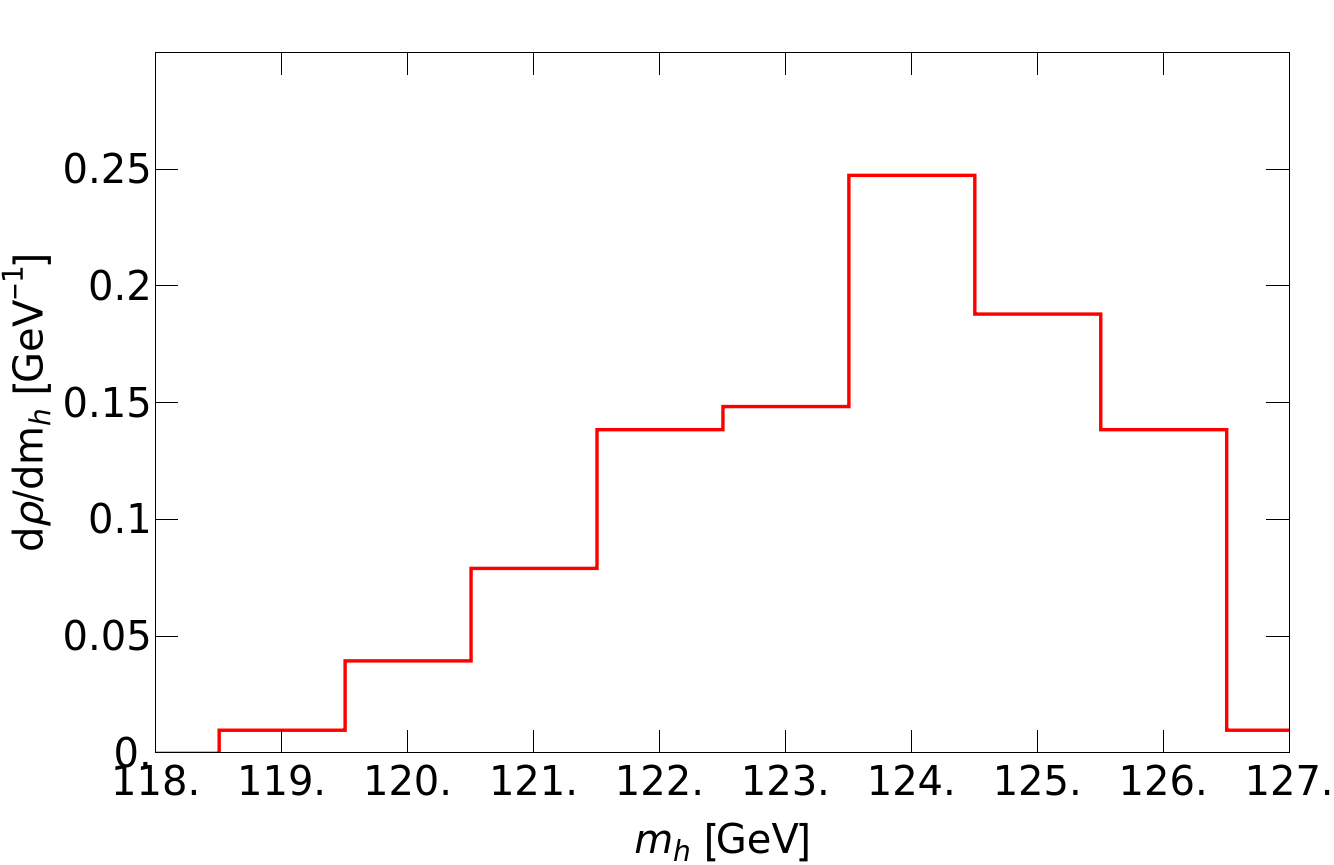}
  \includegraphics[height=0.3\textheight]{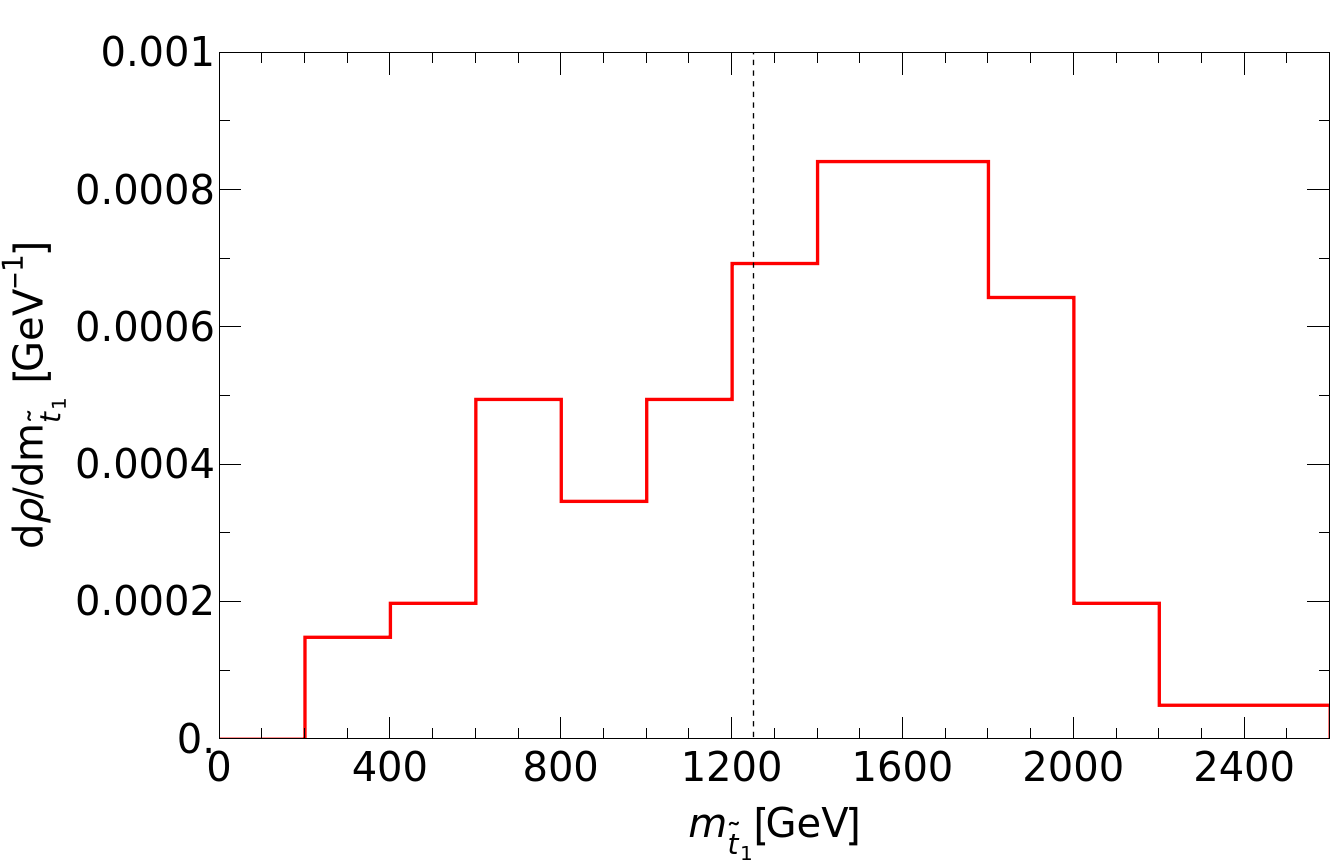}
  \caption{{\it a}) Probability distribution for light Higgs mass $m_h$.
        {\it b}) Probability distribution for lighter top squark
    mass $m_{\tst_1}$.
    We assume statistical selection of soft terms from the string landscape
    with an $n=1$ power-law draw to large soft terms.
  \label{fig:mhmt1}}
\end{center}
\end{figure}

Next, we wish to display properties of top squarks from probability
distributions reflecting stringy naturalness. After our landscape scan,
we display in Fig. \ref{fig:mhmt1} histograms of probability
{\it a}) $dP/dm_h$ and {\it b}) $dP/dm_{\tst_1}$. From frame {\it a}),
we see that the stringy naturalness prefers a light Higgs boson $h$
with mass between $120\ {\rm GeV}<m_h<126$ GeV with a peak at $m_h\sim 125$ GeV. This behavior arises due to maximizing the soft terms that enter into the
radiatively-corrected Higgs mass
\be
m_h^2\simeq m_Z^2\cos^2 2\beta+\frac{3g^2}{8\pi^2}\frac{m_t^4}{m_W^2}\left[\ln
  \frac{m_{\tst}^2}{m_t^2}+\frac{x_t^2}{m_{\tst}^2}\left(1-\frac{x_t^2}{12m_{\tst}^2}\right)\right]
\ee
where $x_t=A_t-\mu\cot\beta$ and $m_{\tst}^2=m_{Q_3}m_{U_3}$ is an effective
stop mass which minimizes log corrections to the scalar potential
(here, $m_{Q_3}$ and $m_{U_3}$ are the third generation doublet and up-squark
soft terms and $A_t$ is the weak scale top-squark trilinear soft term).
For a given value of $m_{\tst}^2$, this expression gives a maximal value for
$m_h$ when $x_t^{max}=\sqrt{6}m_{\tst}$\cite{Carena:2002es,Baer:2011ab,Barger:2012hr}.
The pull on the $A_0$ term to large values (but not so large
as to enter CCB minima in the scalar potential) helps pull $m_h$ up into
the $\sim 125$ GeV range.

In frame {\it b}), we show the probability distribution for $m_{\tst_1}$
from the string landscape with an $n=1$ draw to large soft terms.
While there is just a small probability to have a top squark with mass
below a TeV, the distribution rises to a peak at $m_{\tst_1}\sim 1.5$ TeV
followed by a slow dropoff ending around $m_{\tst_1}\sim 2.5$ TeV.
We also show the present
$m_{\tst_1}\agt 1.2$ TeV limit from ATLAS/CMS searches. By comparing, we see
that LHC experiments are only beginning to probe the range of top squark masses
predicted by the landscape. The reach of LHC Run 3 and HL-LHC should push
into the peak probability region in the coming years, making the search for
light top-squarks of supersymmetry a highly motivated priority.

\begin{figure}[htb!]
\begin{center}
  \includegraphics[height=0.3\textheight]{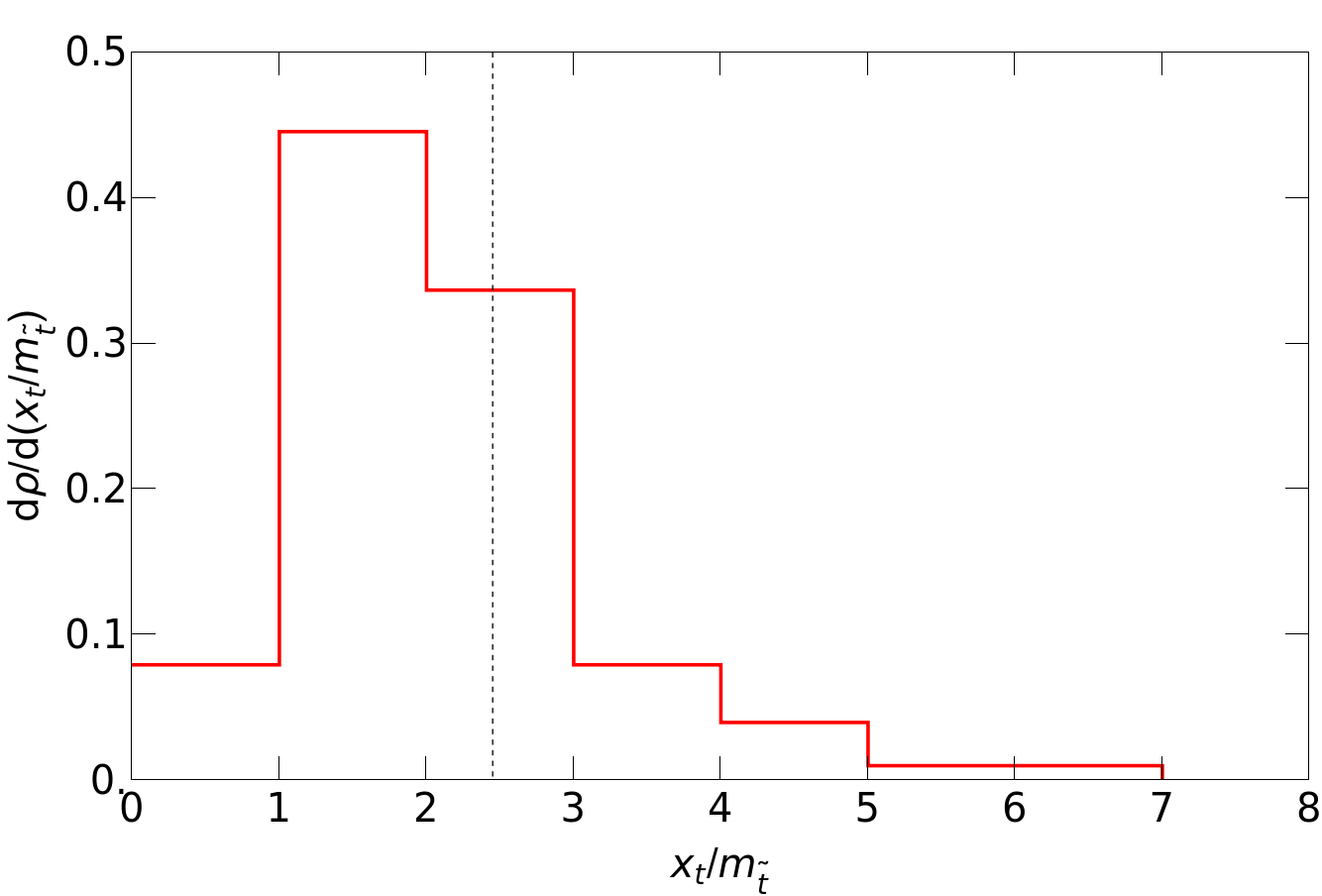}
  \caption{Probability distribution $dP/d(x_t/m_{\tst} )$.
    The vertical dashed line denotes where $x_t=\sqrt{6}m_{\tst}$
    where the Higgs mass radiative corrections becomes maximal.
        We assume statistical selection of soft terms from the string landscape
    with an $n=1$ power-law draw to large soft terms.
  \label{fig:xtmstop}}
\end{center}
\end{figure}

In Fig. \ref{fig:xtmstop}, we show the differential probability distribution
$dP/d(x_t/m_{\tst})$ where $m_{\tst}\equiv\sqrt{m_{\tst_1}m_{\tst_2}}$.
The vertical dashed line denotes where
$x_t=\sqrt{6}m_{\tst}$ which is where the radiative corrections to the Higgs mass
from top squarks are maximal. The distribution peaks just below this point
due to the landscape selection of large trilinear soft terms $A_t$.
This draw to large $A_t$, and hence large $x_t$, helps to understand
why the Higgs mass $m_h$ is pushed up to $m_h\sim 125$ GeV
in the string landscape.

\begin{figure}[htb!]
\begin{center}
  \includegraphics[height=0.3\textheight]{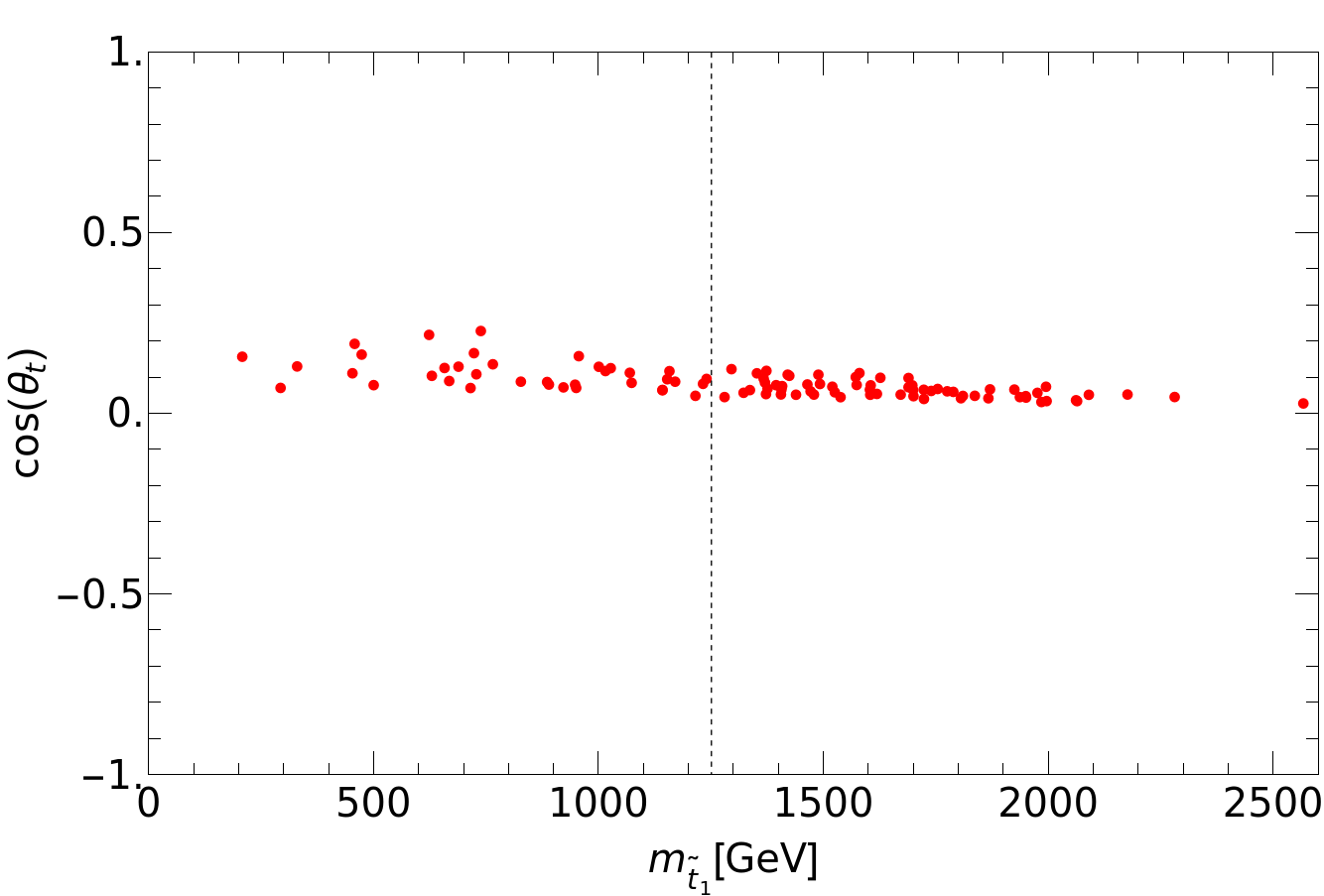}
  \caption{Probability distribution for lighter top squark mass
    vs. $\cos\theta_t$ where $\tst_1= \cos\theta_t\ \tst_L -\sin\theta_t\ \tst_R$.
        We assume statistical selection of soft terms from the string landscape
    with an $n=1$ power-law draw to large soft terms.
  \label{fig:mt1ctht}}
\end{center}
\end{figure}

In Fig. \ref{fig:mt1ctht}, we plot dots of stringy naturalness
in the $m_{\tst_1}$ vs. $\cos\theta_t$ plane where the light top squark
\be
\tst_1=\cot\theta_t\tst_L-\sin\theta_t \tst_R
\ee
in the notation of Ref. \cite{Baer:2006rs} and where $\theta_t$ is the
top squark mixing angle and $\tst_L$ and $\tst_R$ are the weak scale
left- and right-stop eigenstates. From the plot, we see that $\cos\theta_t\sim 0.1$ over the entire expected range of light top squark masses so that we
expect the light top-squark to be predominantly of $\tst_R$ variety in
spite of the expected large stop mixing.
This is because, starting with common soft top-squark masses at the
high scale $m_{Q_3}=m_{U_3}=m_{D_3}\equiv m_0(3)$,
the renormalization group evolution suppresses the right top-squark
soft mass $m_{U_3}$ more than the left top squark soft mass $m_{Q_3}$. 

\subsubsection{$b\to s\gamma$ branching fraction}
\label{ssec:bsg}

A powerful virtual probe of top squark properties comes from
the measured value of the flavor-changing $b$ decay branching fraction
$BR(b\to s\gamma )$. In the SM, this process proceeds via a $tW$ loop
while in 2HDMs there is a comparable contribution from a $tH^\pm$ loop\cite{Barger:1989fj}.
In the MSSM, there are additional contributions from $\tst_i\tchi_{1,2}^\pm$
and even $\tq\tchi_{1,2}^\pm$ loops (the latter tend to decouple in our picture
since first/second generation squarks are drawn to $m_{\tq_i}\sim 10-40$ TeV
level (since their contributions to the weak scale are suppressed by their tiny
Yukawa couplings)). For top-squark and chargino masses nearby to the weak scale,
then the various stop loops tend to dominate the contributions to the
$C_7$ Wilson coefficient albeit with either positive or negative contributions\cite{Baer:1996kv}.
Nonetheless, one expects with rather light top-squarks of a few hundred GeV
that there would be large measured deviations in the
$BR(b\to s\gamma )$ compared to its SM value.
For top-squarks approaching the TeV scale, then these contributions
decouple and one expects the SUSY value for $BR(b\to s\gamma )$ to nearly
match the SM expectation. For our theory calculation, we adopt the
NLO evaluation which is included in Isajet\cite{Baer:1996kv,Baer:1997jq}.
The Isajet value, which doesn't include 2-loop and nonperturbative effects,
asymptotes to $BR(b\to s\gamma )_{SM}^{Isajet}\sim 3.15\times 10^{-4}$.
Thus, along with the Isajet NLO perturbative estimate,
we include a 2-loop and nonperturbative contribution
$\delta\Gamma\equiv \delta\Gamma_{2-loop}+\delta\Gamma_{nonp}\simeq 0.25$
as emphasized by Misiak\cite{Misiak:2018cec}.

The present measured average value from the HFLAV Collaboration\cite{HFLAV:2022pwe}
is given as $BR(b\to s\gamma )=(3.49\pm 0.19)\times 10^{-4}$ which is
dominated by the Belle\cite{Belle:2009nth} and BaBar\cite{BaBar:2012fqh}
measurements. The current SM theory estimate is
$BR(b\to s\gamma )_{SM}^{TH}=(3.36\pm 0.23)\times 10^{-4}$\cite{Misiak:2018cec}.

\begin{figure}[htb!]
\begin{center}
  \includegraphics[height=0.3\textheight]{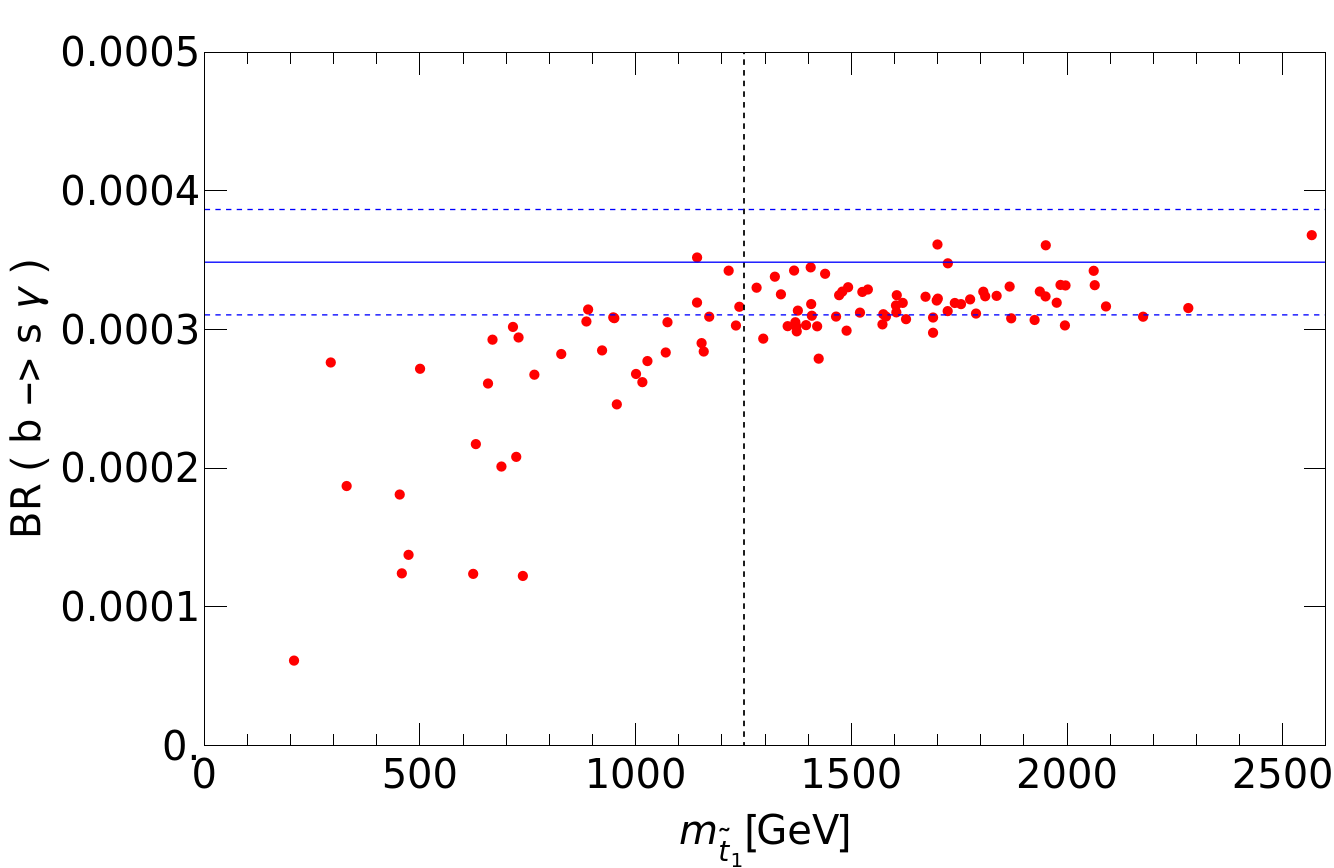}
  \caption{Probability distribution for lighter top squark mass
    vs. $BR(b\to s\gamma )$.
        We assume statistical selection of soft terms from the string landscape
        with an $n=1$ power-law draw to large soft terms.
        The horizontal lines show the PDG measured value $\pm 2\sigma$ error band while the vertical dashed line shows the approximate LHC limit on
        $m_{\tst_1}$ from simplified model analyses.
  \label{fig:mt1bsg}}
\end{center}
\end{figure}

In Fig. \ref{fig:mt1bsg}, we plot points of stringy naturalness in the
$m_{\tst_1}$ vs. $BR(b\to s\gamma )$ plane.
The blue solid line and dashed bands show the HFLAV value $\pm 2\sigma$.
The theory values cluster around $BR(b\to s\gamma )\sim 3.4\times 10^{-4}$
with some larger deviations for lower $m_{\tst_1}\alt 1$ TeV. Thus,
the measured $BR(b\to s\gamma )$ branching fraction tends to support the
scenario of TeV-scale top-squarks as predicted by the string landscape
and as expected from the rather large value of the light Higgs mass
$m_h\sim 125$ GeV.

\subsubsection{Top squark branching fractions}
\label{ssec:BFs}

\begin{figure}[htb!]
\begin{center}
  \includegraphics[height=0.3\textheight]{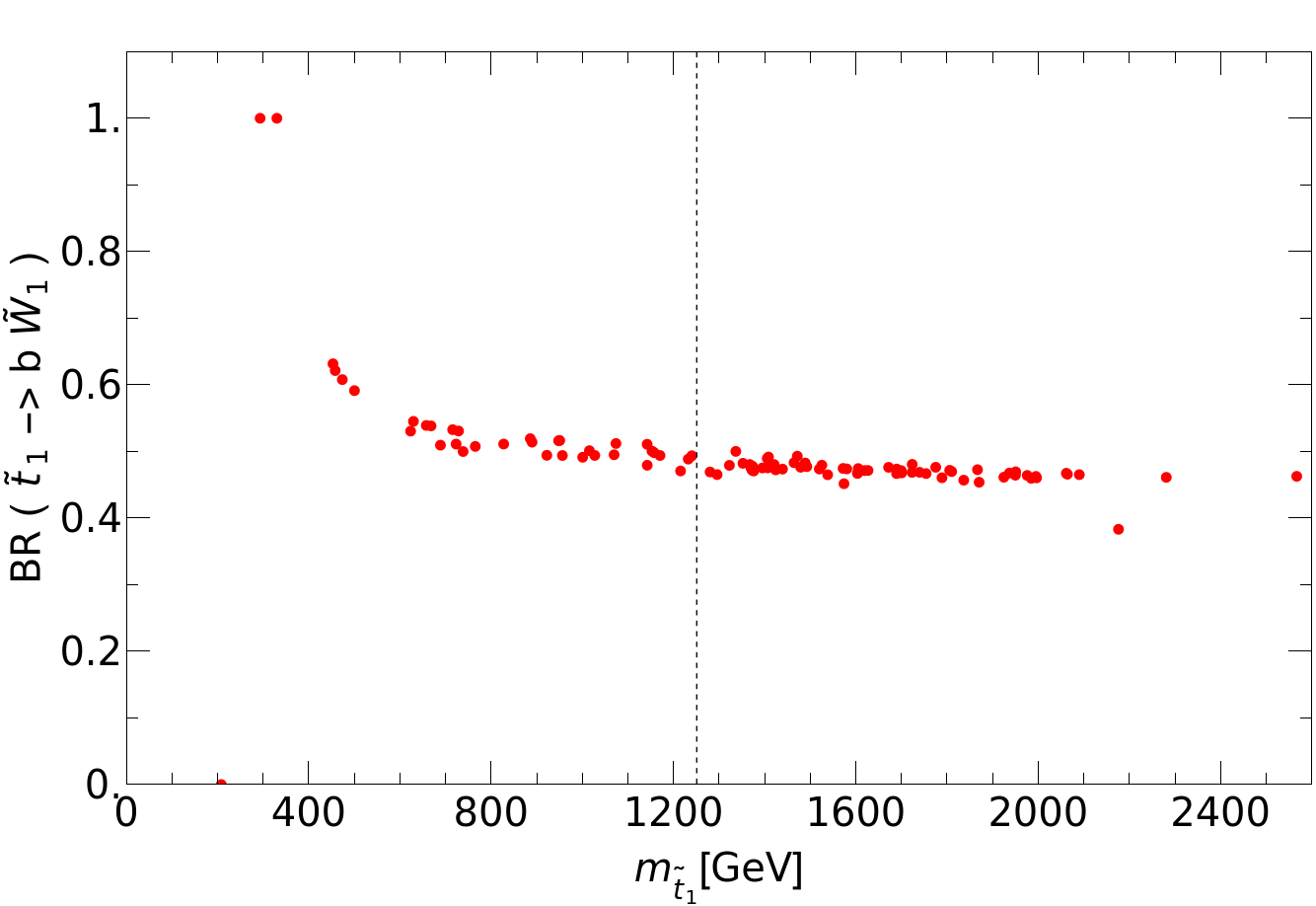}
  \caption{Probability distribution for lighter top squark mass
    vs. $BF(\tst_1\to b\tchi_1^+ )$.
        We assume statistical selection of soft terms from the string landscape
    with an $n=1$ power-law draw to large soft terms.
  \label{fig:mt1BFt1bw1}}
\end{center}
\end{figure}

The top squark decay widths $\Gamma (\tst_1\to t\tchi_i^0)$ ($i=1-4$) and
$\Gamma (\tst_1\to b\tchi_j^+$) ($j=1-2$) are expected to be the dominant
top-squark decay modes and their formulae are listed in Ref. \cite{Baer:2006rs}
as Equations {\it B.39} and {\it B.43} respectively.
The numerical values can be extracted from the Isajet\cite{Paige:2003mg} code.
The decay widths depend sensitively on the top-squark gauge couplings
and the top-quark Yukawa coupling along with the mixing angle $\theta_t$
and the decay kinematics.
In Fig. \ref{fig:mt1BFt1bw1} we show the stringy natural values of
$BF(\tst_1\to b\tchi_1^+ )$ vs. $m_{\tst_1}$. From the plot, we see a rather
uniform prediction vs. $m_{\tst_1}$ that $BF(\tst_1\to b\tchi_1^+ )$ occurs
very close to the 50\% level. 

\begin{figure}[htb!]
\begin{center}
  \includegraphics[height=0.3\textheight]{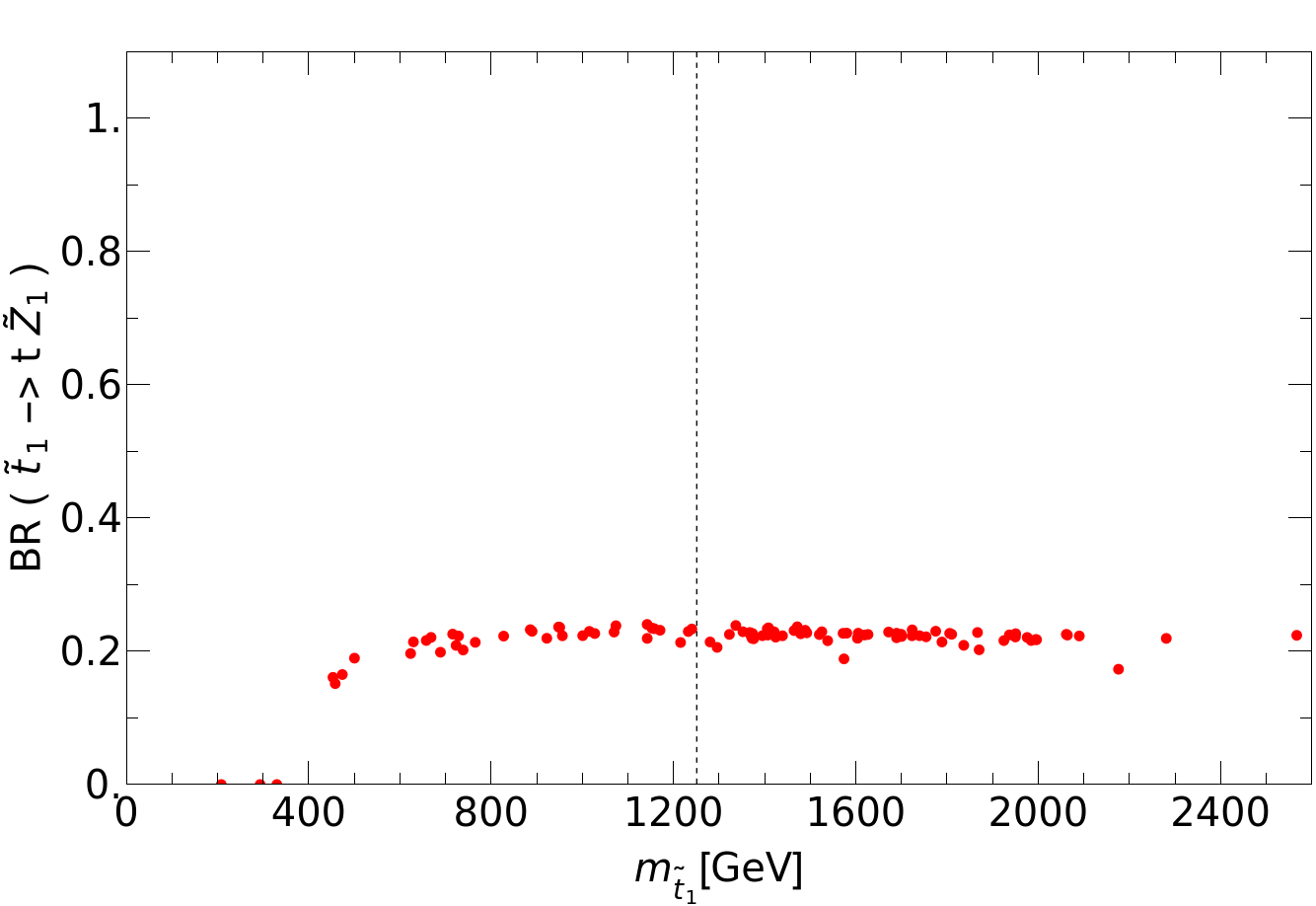}
  \caption{Probability distribution for lighter top squark mass
    vs. $BF(\tst_1\to t\tchi_1^0 )$.
        We assume statistical selection of soft terms from the string landscape
    with an $n=1$ power-law draw to large soft terms.
  \label{fig:mt1BFt1tz1}}
\end{center}
\end{figure}

\begin{figure}[htb!]
\begin{center}
  \includegraphics[height=0.3\textheight]{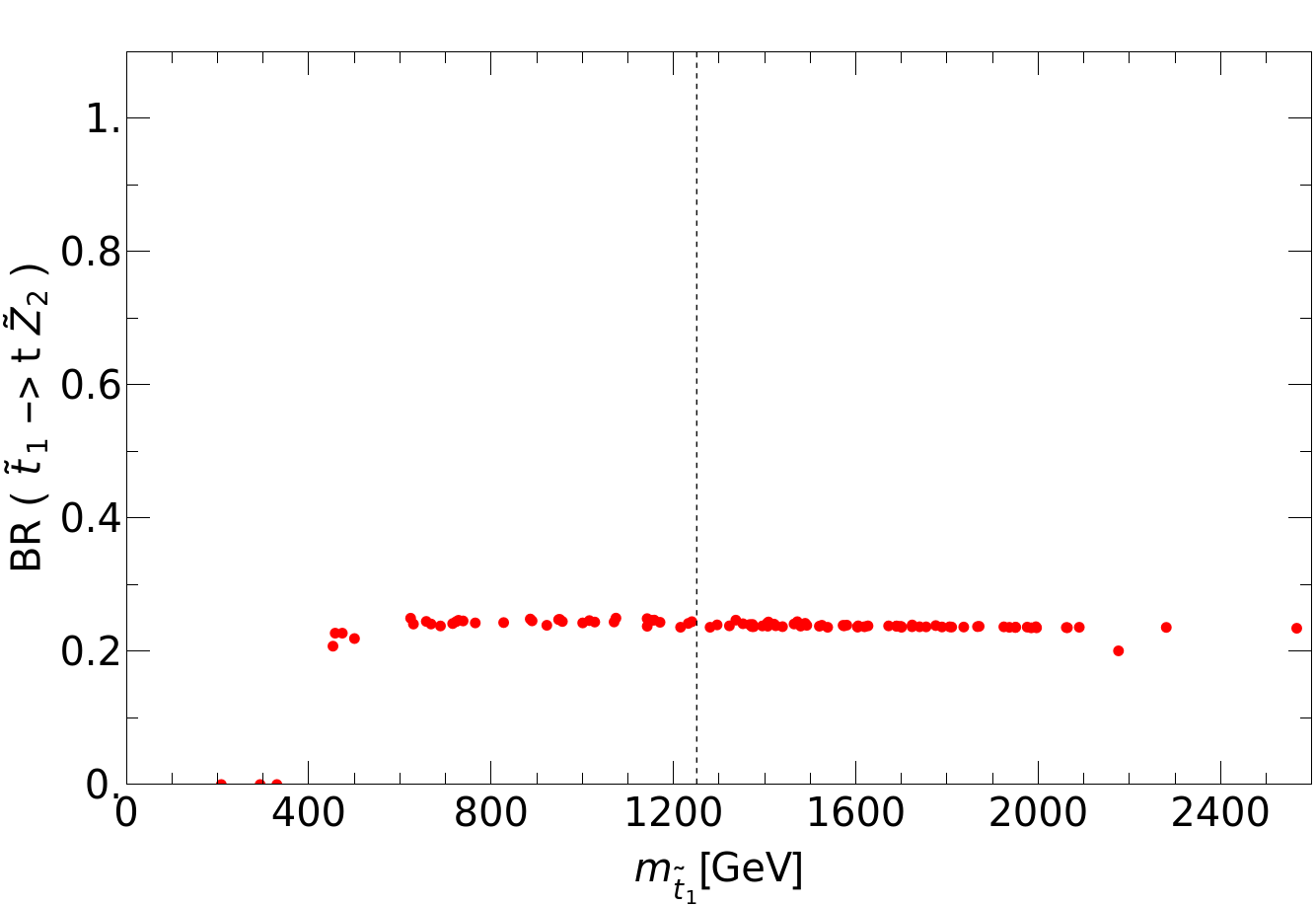}
  \caption{Probability distribution for lighter top squark mass
    vs. $BF(\tst_1\to t\tchi_2^0 )$.
        We assume statistical selection of soft terms from the string landscape
    with an $n=1$ power-law draw to large soft terms.
  \label{fig:mt1BFt1tz2}}
\end{center}
\end{figure}

In Fig. \ref{fig:mt1BFt1tz1}, we show the prediction for
$BF(\tst_1\to t\tchi_1^0 )$ vs. $m_{\tst_1}$. The result here is also
rather uniform in $m_{\tst_1}$: that $BF(\tst_1\to t\tchi_1^0 )\sim 20-25\%$.
Likewise, in Fig. \ref{fig:mt1BFt1tz2} we show the $BF(\tst_1\to t\tchi_2^0)$.
This branching fraction also tends to occur at the 20-25\% level with little
variation vs. $m_{\tst_1}$. Further branching fractions such as
$BF(\tst_1\to t\tchi_3^0 )$ can occur at the several percent level,
while others such as $\tst_1\to t\tchi_4^0$ and $\tst_1\to b\tchi_2^+$
tend to occur at the sub-percent level.

\section{Production and decay of top squarks at LHC}
\label{sec:LHC}

\begin{figure}[htb!]
\begin{center}
  \includegraphics[height=0.3\textheight]{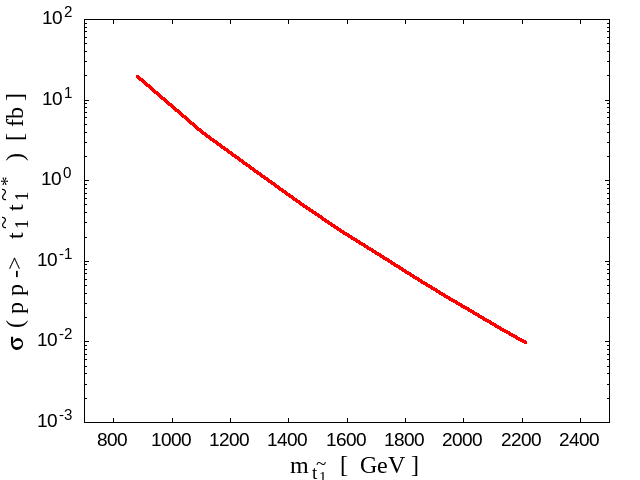}
  \caption{Plot of $\sigma (pp\to \tst_1\tst_1^*X)$ from Prospino (NLO) versus
    $m_{\tst_1}$ for $pp$ collisions at $\sqrt{s}=14$ TeV.
\label{fig:sigt1t1}}
\end{center}
\end{figure}

For the benefit of the reader, we show in Fig. \ref{fig:sigt1t1}
the next-to-leading-order (NLO) Prospino\cite{Beenakker:1996ed}
prediction for top squark pair production at LHC with $\sqrt{s}=14$ TeV
collisions: $\sigma (pp\to\tst_1\tst_1^* X )$ vs. $m_{\tst_1}$.
Starting just above the present LHC excluded region,
with $m_{\tst_1}=1.25$ TeV, we find $\sigma (\tst_1\tst_1^*)\sim 1$ fb,
corresponding to 3000 signal events assuming the nominal HL-LHC
integrated luminosity of 3 ab$^{-1}$. Even for $m_{\tst_1}$ as high as 2 TeV,
we find $\sigma (\tst_1\tst_1^*)\sim 0.02$ fb, corresponding to 60 signal
events at HL-LHC before cuts.

\begin{figure}[htb!]
\begin{center}
  \includegraphics[height=0.3\textheight]{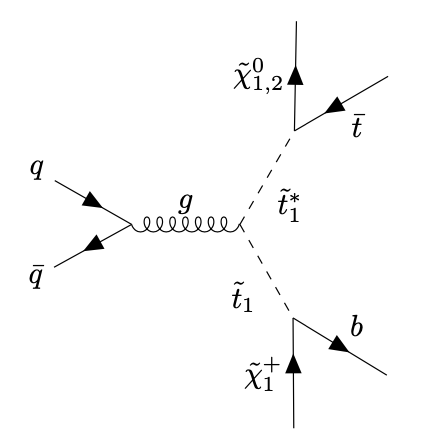}
  \caption{Representative diagram for top squark pair production and decay
    at LHC in natural SUSY.
\label{fig:diagram}}
\end{center}
\end{figure}

The projected HL-LHC reach for top squark pair producton is usually
presented in terms of simplified models by the ATLAS and CMS collaborations
where a single top squark decay mode is assumed.
We see from the previous subsection that such analyses are not realistic
from the point of view of the string landscape and so we will examine the
reach of LHC at HL-LHC using the several predicted decay modes.
This will give rise to mixed decay mode configurations such as is shown
in Fig. \ref{fig:diagram} where one $\tst_1\to b\tchi_1^+$ and the other
$\tst_1\to t\tchi_1^0$.
Thus, we expect three main signal channels:
\bi
\item $b\bar{b}+\eslt$,
\item $t\bar{t}+\eslt$ and 
\item $t\bar{b}+\eslt$ (plus charge conjugate mode).
\ei
In addition, some subset of events will contain soft decay products
from the unstable higgsinos in the cascade decay. Of particular note
is the ocassional presence of $\tchi_2^0\to\tchi_1^0\ell\bar{\ell}$
(with $\ell = e$ or $\mu$) where $m(\ell\bar{\ell})<m_{\tchi_2^0}-m_{\tchi_1^0}$.

\section{A stringy natural top squark benchmark point}
\label{sec:BM}

In the following Section, we will examine top squark pair production
at LHC14 for the following {\it Benchmark Point} (BM)  which is typical
of stringy natural models.
The BM point comes from the NUHM2 model\cite{Baer:2005bu}
with parameters as listed in the Table~\ref{tab:bm}. It is a natural SUSY benchmark point
since $\Delta_{EW}=22$, even though the lightest top squark lies at
$m_{\tst_1}\sim 1.7$ TeV and $m_{\tg}\sim 2.8$ TeV.
The spectra is generated using the Isasugra code\cite{Baer:1994nc}
from Isajet\cite{Paige:2003mg}.
We can also expand this natural SUSY BM point into a natural SUSY
{\it Model Line} by simply varying the $A_0$ parameter which results in
variation of $m_{\tst_1}:800-2200$ GeV while hardly changing $m_h$ or other
sparticle masses. 
\begin{table}[h!]
\centering
\begin{tabular}{lc}
\hline
parameter & stringy natural BM point \\
\hline
$m_0$      & 5 TeV \\
$m_{1/2}$      & 1.2 TeV \\
$A_0$      & -8 TeV \\
$\tan\beta$    & 10  \\
\hline
$\mu$          & 250 GeV  \\
$m_A$          & 2 TeV \\
\hline
$m_{\tilde{g}}$   & 2830 GeV \\
$m_{\tilde{u}_L}$ & 5440 GeV \\
$m_{\tilde{u}_R}$ & 5561 GeV \\
$m_{\tilde{e}_R}$ & 4822 GeV \\
$m_{\tilde{t}_1}$& 1714 GeV \\
$m_{\tilde{t}_2}$& 3915 GeV \\
$m_{\tilde{b}_1}$ & 3949 GeV \\
$m_{\tilde{b}_2}$ & 5287 GeV \\
$m_{\tilde{\tau}_1}$ & 4746 GeV \\
$m_{\tilde{\tau}_2}$ & 5110 GeV \\
$m_{\tilde{\nu}_{\tau}}$ & 5107 GeV \\
$m_{\tilde{\chi}_1^\pm}$ & 261.7 GeV \\
$m_{\tilde{\chi}_2^\pm}$ & 1020.6 GeV \\
$m_{\tilde{\chi}_1^0}$ & 248.1 GeV \\ 
$m_{\tilde{\chi}_2^0}$ & 259.2 GeV \\ 
$m_{\tilde{\chi}_3^0}$ & 541.0 GeV \\ 
$m_{\tilde{\chi}_4^0}$ & 1033.9 GeV \\ 
$m_h$       & 124.7 GeV \\ 
\hline
$\Omega_{\tilde{\chi}_1}^{std}h^2$ & 0.016 \\
$BR(b\to s\gamma)\times 10^4$ & $3.1$ \\
$BR(B_s\to \mu^+\mu^-)\times 10^9$ & $3.8$ \\
$\sigma^{SI}(\tilde{\chi}_1^0, p)$ (pb) & $2.2\times 10^{-9}$ \\
$\sigma^{SD}(\tilde{\chi}_1^0, p)$ (pb)  & $2.9\times 10^{-5}$ \\
$\langle\sigma v\rangle |_{v\to 0}$  (cm$^3$/sec)  & $1.3\times 10^{-25}$ \\
$\Delta_{\rm EW}$ & 22 \\
\hline
\end{tabular}
\caption{Input parameters (TeV) and masses (GeV)
for the stringy natural SUSY benchmark point from the NUHM2 model
with $m_t=173.2$ GeV using Isajet 7.88~\cite{Paige:2003mg}.
}
\label{tab:bm}
\end{table}

\section{Reach of LHC for natural top squarks}
\label{sec:reach}

We next examine the reach of HL-LHC ($\sqrt{s}=14$ TeV with 3000 fb$^{-1}$)
for the top squarks of stringy natural SUSY.
To proceed, we generate a SUSY Les Houches Accord (SLHA) file\cite{Skands:2003cj} for our
natural SUSY BM point and feed this into Pythia\cite{Sjostrand:2006za}
which is used for signal and the $2\to 2$ background (BG) processes.
For the $2\to 3$ BG processes, we use Madgraph\cite{Alwall:2011uj}
coupled to Pythia.
The SM BGs considered are: $t\bar{t}$, $b\bar{b}Z$, $t\bar{t}Z$, $t\bar{t}W$,
$b\bar{b}W$ and single-top production.
We adopt the toy detector simulation Delphes\cite{deFavereau:2013fsa}.

The baseline reconstructed objects are as follows.

{\it Baseline small radius(SR) jet}:
\begin{enumerate}
\item Found by anti-$k_t$ jet finder algorithm with $p_T(min)= 25$ GeV and
  $R = 0.4$ and
\item $|\eta (j)| < 4.5$.
\end{enumerate}
{\it Isolated lepton}:
\begin{enumerate}
\item $|\eta (\mu ) | < 2.5$ for muon, $|\eta (e ) |< 2.47$ for electron,
\item $p_T(\mu) > 25$ GeV for muon, $p_T(e) > 20$ GeV for electron.
\end{enumerate}
{\it Large radius(LR) jet}:
\begin{enumerate}
\item Found by Cambridge/Aachen finder algorithm with $p_T(min)= 400$ GeV
  with $R = 1.5$.
\end{enumerate}
For signal objects, we also require
{\it signal b-jets}:
\begin{enumerate}
\item satisfy the baseline SR jet requirement above.
\item $|\eta (b)| < 2.4$ and
\item tagged by Delphes as $b$-jet.
\end{enumerate}
The {\it signal top candidate} is reconstructed with either of the
following criteria:
\begin{itemize}
\item The fat jet $J$ is tagged by the HEPTopTagger2\cite{Anders:2013oga,Plehn:2011sj} as a top.
  In such case, the top 4-vector reconstructed by the tagger is used for
  further kinematics calculations, or
\item The fat jet $J$ has a trimmed mass $115\ {\rm GeV} < m_J < 225$ GeV
  and has at least 1 $b$-jet within the cone radius of the fat jet
  $(\Delta R(J,b) < 1.5$). In such a case, the trimmed 4-vector of the fat jet
  is used for further kinematics calculations.
\end{itemize}

The events then are separated into three channels: $t\bar{t}+\eslt$, $tb+\eslt$, and $b\bar{b}+\eslt$.
The workflow to determine each channel is as follows:
\begin{itemize}
\item If there are at least 2 tops being tagged by the HEPTopTagger2,
the two tops with the hardest $p_T$ are chosen as signals,
and this channel is labeled as $t\bar{t}+\eslt$.
\item Otherwise, if the HEPTopTagger2 tags 1 top, and the trimming method
  tags at least one other, this channel is labeled again as $t\bar{t}+\eslt$.
  The top tagged by the HEPTopTagger2, and the hardest fat jet found by the
  trimming method are chosen as signals.
\item Otherwise, if the HEPTopTagger2 fails to tag any tops,
  but the trimming method found at least two, this channel is also labeled
  as $t\bar{t}+\eslt$. The two tops with the hardest $p_T$ are chosen as signals.
\item Otherwise, if there is exactly 1 top tagged by either HEPTopTagger2
  or the trimming method, then look for extra $b$-jet candidates.
  The $b$-jet candidates must satisfy signal $b$-jet requirement listed above.
  The $b$-jet candidate needs to be well separated with the 3 subjets of the
  reconstructed top (Both HEPTopTagger2 and the trimming algorithm can
  provide the subjet 4-vectors): $\Delta R(subjet, b) > 0.4$.
  If there are $b$-jets satisfying these requirements, the $b$ candidate
  that minimizes the vector sum of $\eslt+p_T(t)+p_T(b)$ is chosen as the signal. This channel is labeled as $tb+\eslt$.
\item Otherwise, if the event fails any of the above selection requirements
  but has at least two $b$ jets satisfied the signal $b$-jet requirement,
  this channel is labeled as $b\bar{b}+\eslt$.
  The pair of $b$-jets that minimize the vector sum of $\eslt+p_T(b_1)+p_T(b_2)$
  are chosen as signals. The harder of the two is labeled as $b_1$,
  and the other is $b_2$ in the following.
\end{itemize}

\subsection{$b\bar{b}+\eslt$}
\label{ssec:bbMET}

We first examine the $b\bar{b}+\eslt$ channel.
After examining various distributions, we require the following.
\bi
\item $\eslt > 800$ GeV,
\item $|\eta (b_1)| < 2.0$,
\item $p_T(b_1) > 200$ GeV,
\item $H_T > 1500$ GeV,
\item $min[m_T(b_1, \eslt), m_T(b_2, \eslt)] > 175$ GeV,
\item $min(\Delta \phi(b, \eslt)) > 20^\circ$, where $b$ loops over all $b$-jets
  in the event.
\ei

\begin{figure}[htb!]
\begin{center}
  \includegraphics[height=0.4\textheight]{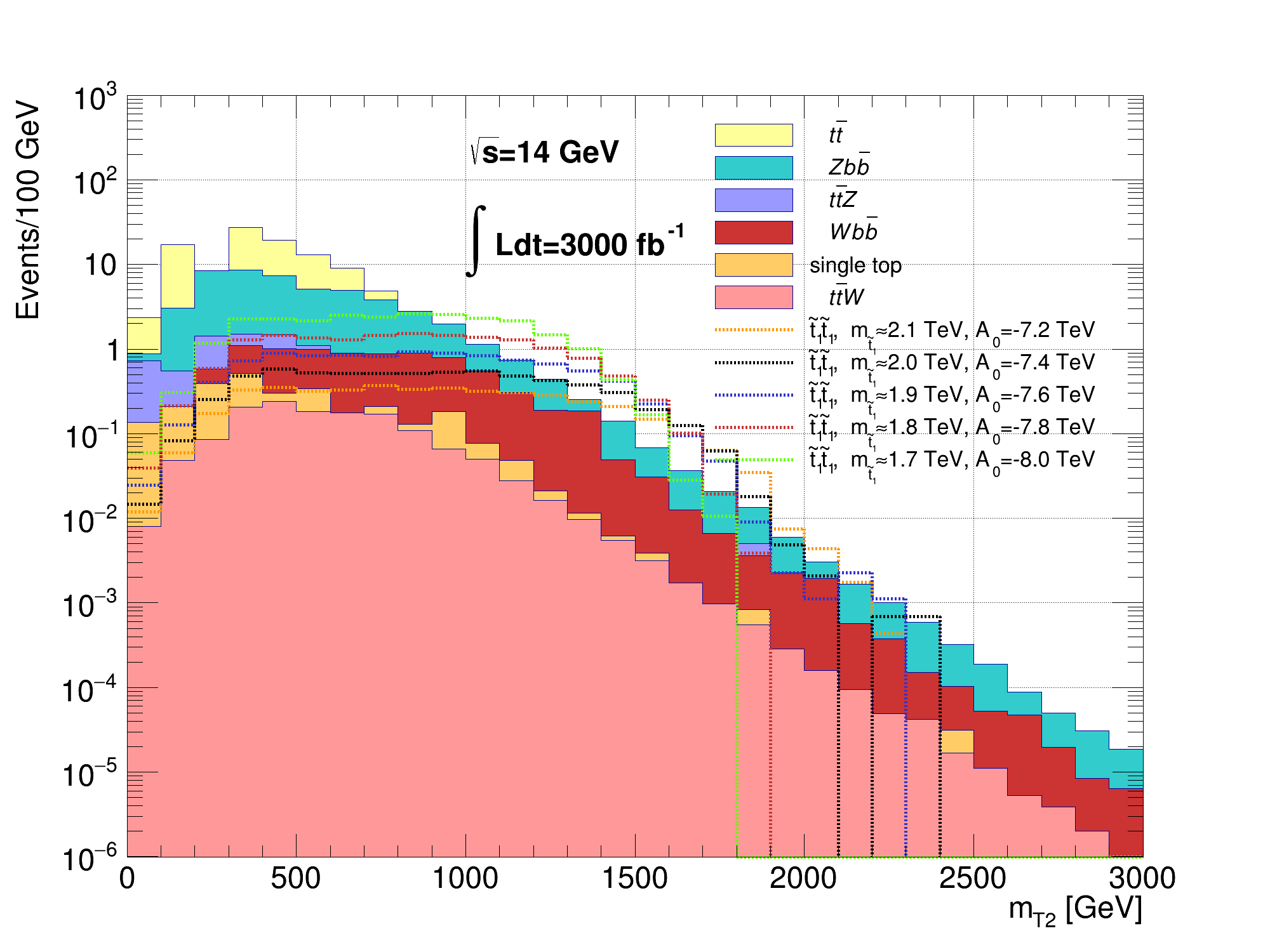}
  \caption{Distribution in $m_{T2}$ from top-squark pair production
    at LHC14 in the $b\Bar{b}+\eslt$ channel along with dominant SM backgrounds
    after cuts listed in the text.
\label{fig:mt2_2b}}
\end{center}
\end{figure}

After these cuts, we construct the $m_{T_2}$ distribution\cite{Barr:2003rg}
and plot the resultant distribution in Fig. \ref{fig:mt2_2b}.
The strategy becomes clear: look for a high $m_{T_2}$ deviation from
expected background at the higher $m_{T_2}$ values where we expect
$m_{T_2}$ to be bounded from above by $m_{\tst_1}$.
In the plot, we show signal histograms for five different values of
$m_{\tst_1}$ along with leading backgrounds. While BG does indeed dominate at
low $m_{T_2}$, a signal emerges from BG at higher values.
If there are sufficient number of signal events above BG, then a signal can be
claimed.

\subsection{$tb +\eslt$}
\label{ssec:tbMET}

After examining various distributions, for the $tb+\eslt$ channel we require
\bi
\item $\eslt > 400$ GeV,
\item $H_T > 1400$ GeV,
\item $L_T > 1800$ GeV (defined as the scalar sum of $p_T(t) + p_T(b) + \eslt$),
  where $t$ and $b$ here are the signal top and $b$-jet.
\item $min[m_T(t, \eslt), m_T(b, \eslt)] > 175$ GeV, 
\item $min(\Delta\phi(b, \eslt)) > 40^\circ$, where $b$ loops over all $b$-jets
  in the event.
\item $min(\Delta\phi (J, \eslt)) > 30^\circ$, where $J$ loops over all
  fat jets in the event, no matter whether they've been tagged as
  top or not.
  \ei

\begin{figure}[htb!]
\begin{center}
  \includegraphics[height=0.4\textheight]{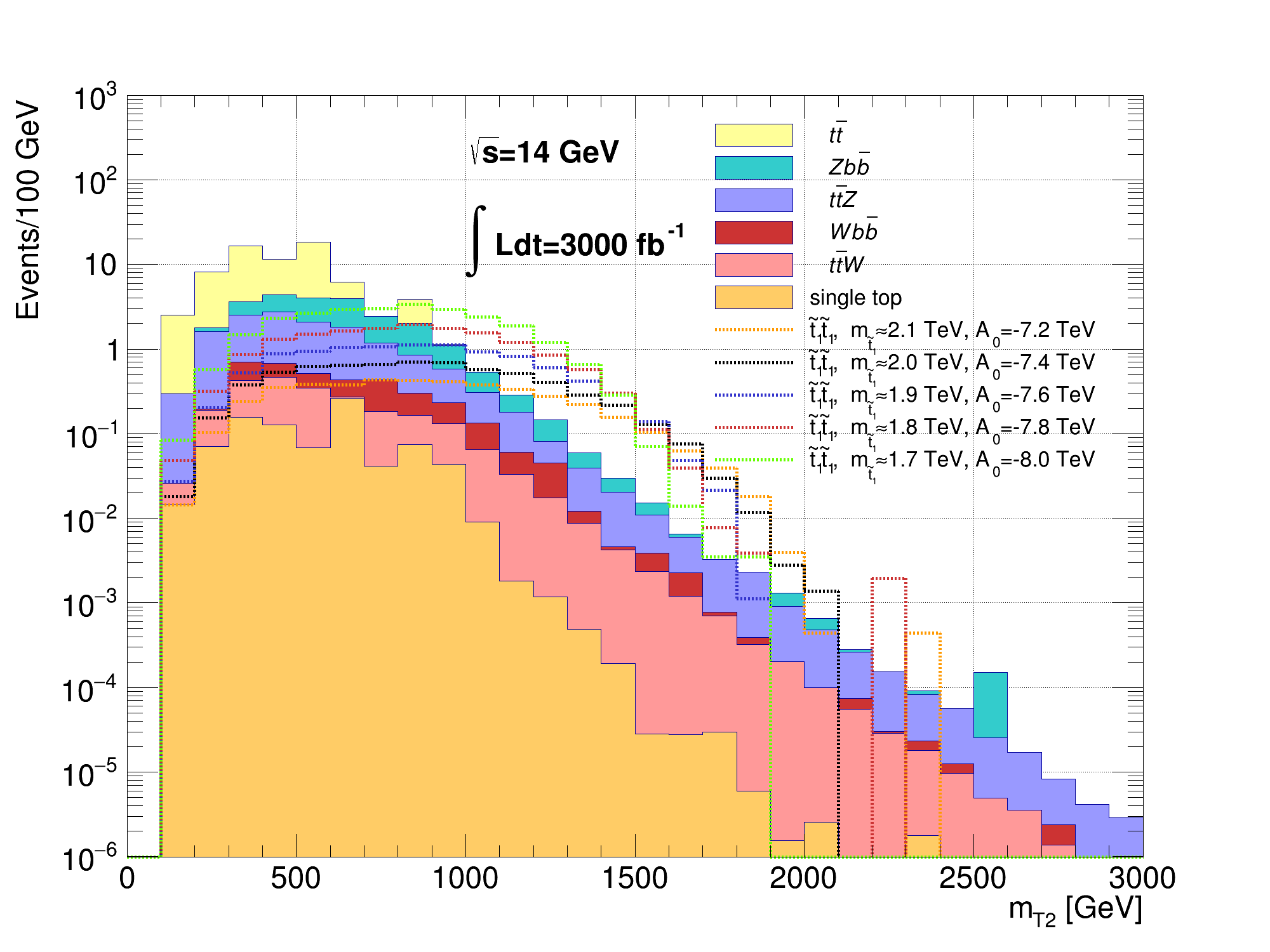}
  \caption{Distribution in $m_{T2}$ from top-squark pair production
    at LHC14 in the $tb+\eslt$ channel along with dominant SM backgrounds
    after cuts listed in the text.
\label{fig:mt2_bt}}
\end{center}
\end{figure}

  The resultant $m_{T_2}$ distribution is shown in Fig. \ref{fig:mt2_bt} where
  again we expect the signal distribution to be bounded from above by
  $m_{\tst_1}$ whilst BG is a continuum. The five signal histograms do indeed
  emerge from BG at high $m_{T_2}$ although not necessarily at an observable rate.
  The largest BG at high $m_{T_2}$ is from $t\bar{t}Z$ production.

\subsection{$t\bar{t}+\eslt$}
\label{ssec:ttMET}

Next, we examine various distributions for signal and BG in
the $t\bar{t}+\eslt$ signal channel.
We then require the following:
\bi
\item $\eslt > 300$ GeV,
\item $H_T > 1400$ GeV,
\item $min[m_T(t_1, \eslt), m_T(t_2, \eslt)] > 175$ GeV,
\item $min(\Delta\phi (b, \eslt)) > 40^\circ$, where $b$ loops over all $b$-jets
  in the event.
\item $min(\Delta\phi (J, \eslt)) > 30^\circ$, where $J$ loops over all fat
  jets in the event, no matter whether they've been tagged as top or not.
  \ei

\begin{figure}[htb!]
\begin{center}
  \includegraphics[height=0.4\textheight]{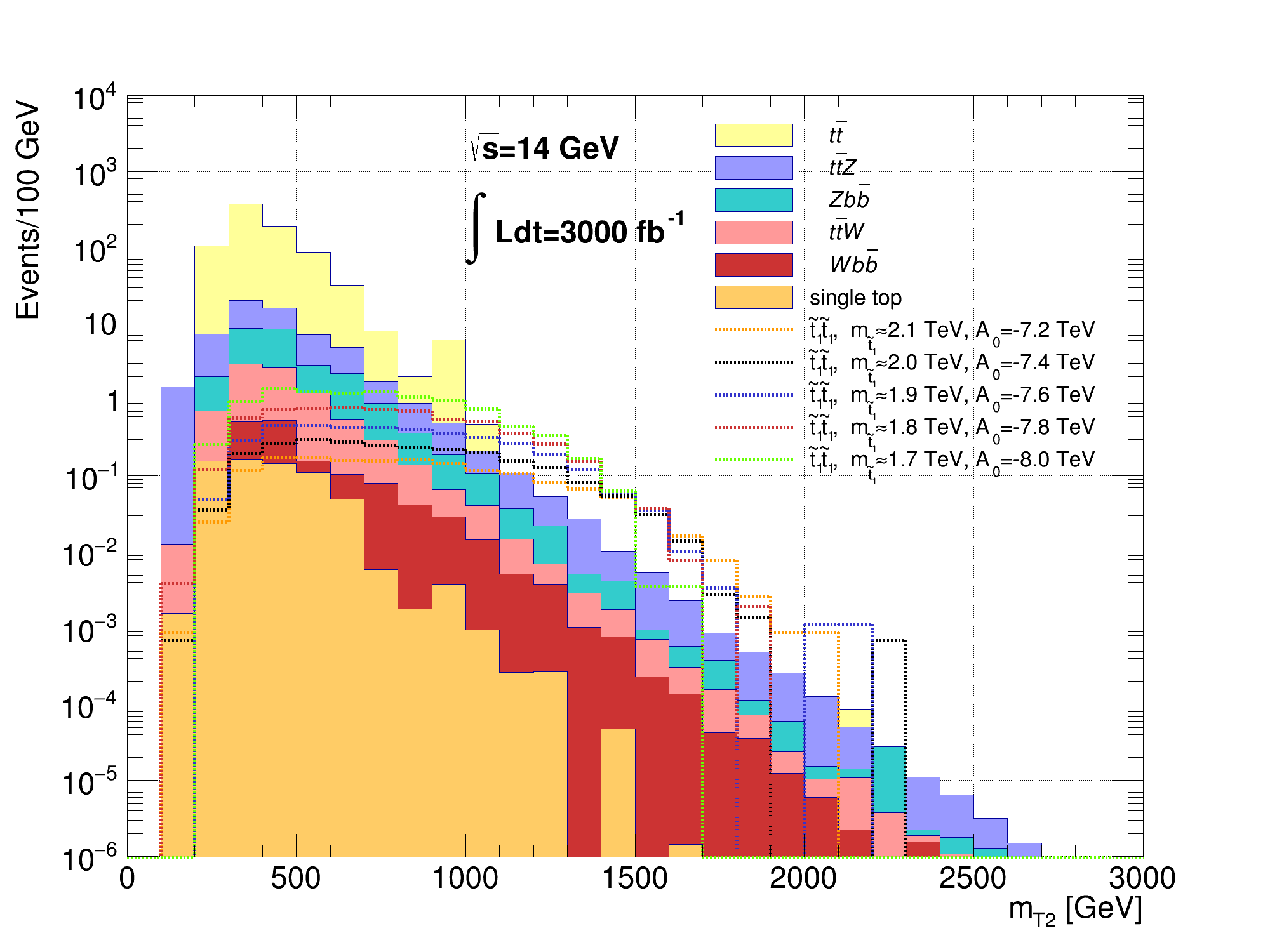}
  \caption{Distribution in $m_{T2}$ from top-squark pair production
    at LHC14 in the $t\Bar{t}+\eslt$ channel along with dominant SM backgrounds
    after cuts listed in the text.
\label{fig:mt2_2t}}
\end{center}
\end{figure}

  The subsequent $m_{T_2}$ distribution is plotted in Fig. \ref{fig:mt2_2t}.
  The signal distributions emerge from SM BG at high $m_{T_2}$ but at more
  marginal rates than the other channels due to the lower efficiency
  to tag top-jets.

\subsection{Cumulative reach of HL-LHC for top squark pair production in natSUSY}
\label{ssec:reach}

Using the analysis cuts for the various signal channels discussed
above, we can now create reach plots to show the HL-LHC discovery sensitivity
versus $m_{\tst_1}$ along our natural SUSY model line.
We use the $5\sigma$ level to claim discovery of a 
top-squark and assume the true distribution one observes
experimentally corresponds to signal-plus-background.
We then test this against the background-only distribution in order to
see if the background-only hypothesis can be rejected at the $5\sigma$ level.
Specifically, we use the binned $m_{T_2}$ distributions (bin width
of 100 GeV) from each signal channel as displayed above to obtain the
discovery/exclusion limits.

In the case of the exclusion line, the upper limits for exclusion
of a signal are set at 95\% CL; one assumes the true distribution
one observes in experiment corresponds to background-only.
The limits are then computed using a modified frequentist $CL_s$
method\cite{Read_2002} where the profile likelihood ratio is
the test statistic. 
For both the exclusion and discovery plots, the asymptotic approximation
for obtaining the median significance is employed\cite{Cowan_2011}.
For both discovery and exclusion estimates, we combine results from
all three top-squark signal channels: $t\bar{t}+\eslt$, $b\bar{b}+\eslt$ and $tb+\eslt$.

\begin{figure}[htb!]
\begin{center}
  \includegraphics[height=0.4\textheight]{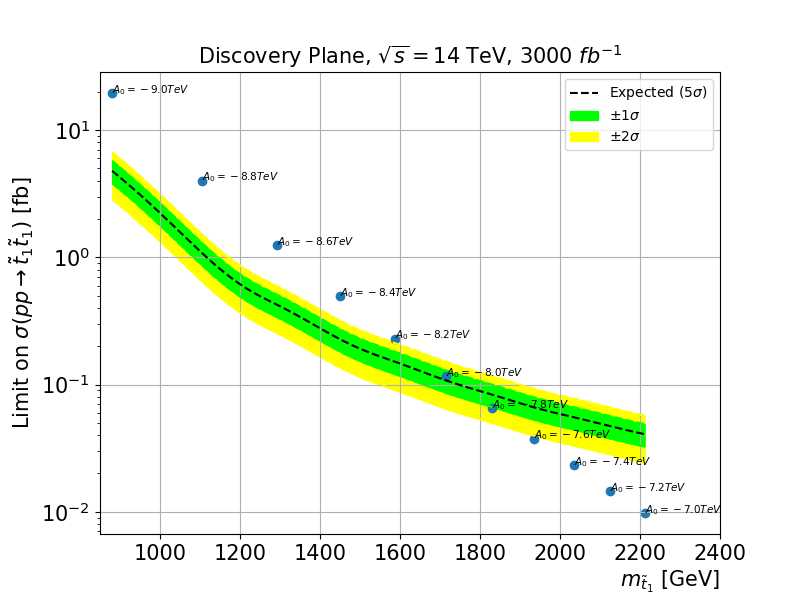}\\
    \includegraphics[height=0.4\textheight]{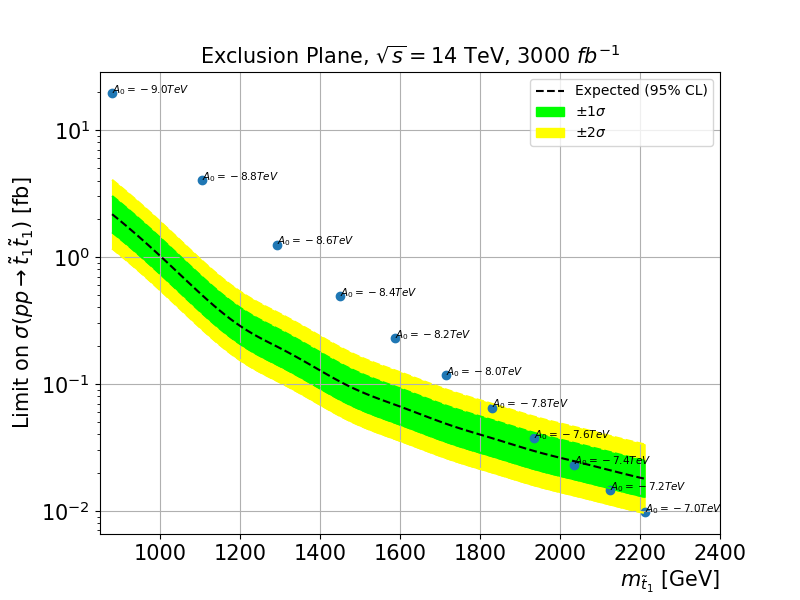}\\
  \caption{Expected 5$\sigma$ discovery limit and expected 95$\%$ CL exclusion limit on top-squark pair production cross section 
    vs. $m_{\tst_1}$ from a natural SUSY model line at HL-LHC with
    $\sqrt{s}=14$ TeV and 3000 fb$^{-1}$ of integrated luminosity.
\label{fig:discexcl}}
\end{center}
\end{figure}

In Fig. \ref{fig:discexcl}{\it a}), we show the $5\sigma$ discovery
cross section as the dashed line along with 1- and 2-$\sigma$
error bands. We show the corresponding natural SUSY model line as blue dots.
We see that HL-LHC with 3000 fb$^{-1}$ fb of integrated luminosity can
discover natural SUSY top-squarks out to $m_{\tst_1}\sim 1700$ GeV.
In Fig. \ref{fig:discexcl}{\it b}), we plot the HL-LHC 95\% exclusion reach.
In this case, the exclusion reach extends out to $m_{\tst_1}\sim 2000$ GeV.
By comparing these results with expectations from stringy naturalness
in Fig. \ref{fig:mhmt1}{\it b}),
we see that HL-LHC can cover the bulk of stringy natural
parameter space, although a tail of probability does extend past
$m_{\tst_1}\sim 2$ TeV.

\section{Conclusions}
\label{sec:conclude}

In this paper we have examined what sort of values of top squark masses
and other properties are expected from the string landscape where a
power-law draw to large soft terms is expected, but where the derived
value of the weak scale must lie within the ABDS window in order to
allow for complex nuclei (and hence atoms) in each anthropically-allowed
pocket universe. Under this {\it stringy naturalness} requirement,
we find $m_{\tst_1}\sim 1-2.5$ TeV with large mixing.
These results are in accord with measurements of $BR(b\to s\gamma )$ which
are suggestive of TeV-scale top-squarks so that SUSY contributions
to this decay rate decouple.
The large mixing helps boost $m_h\to 125$ GeV while minimizing the top
squark contributions to the weak scale $\Sigma_u^u(\tst_{1,2})$.

\begin{figure}[htb!]
\begin{center}
  \includegraphics[height=0.3\textheight]{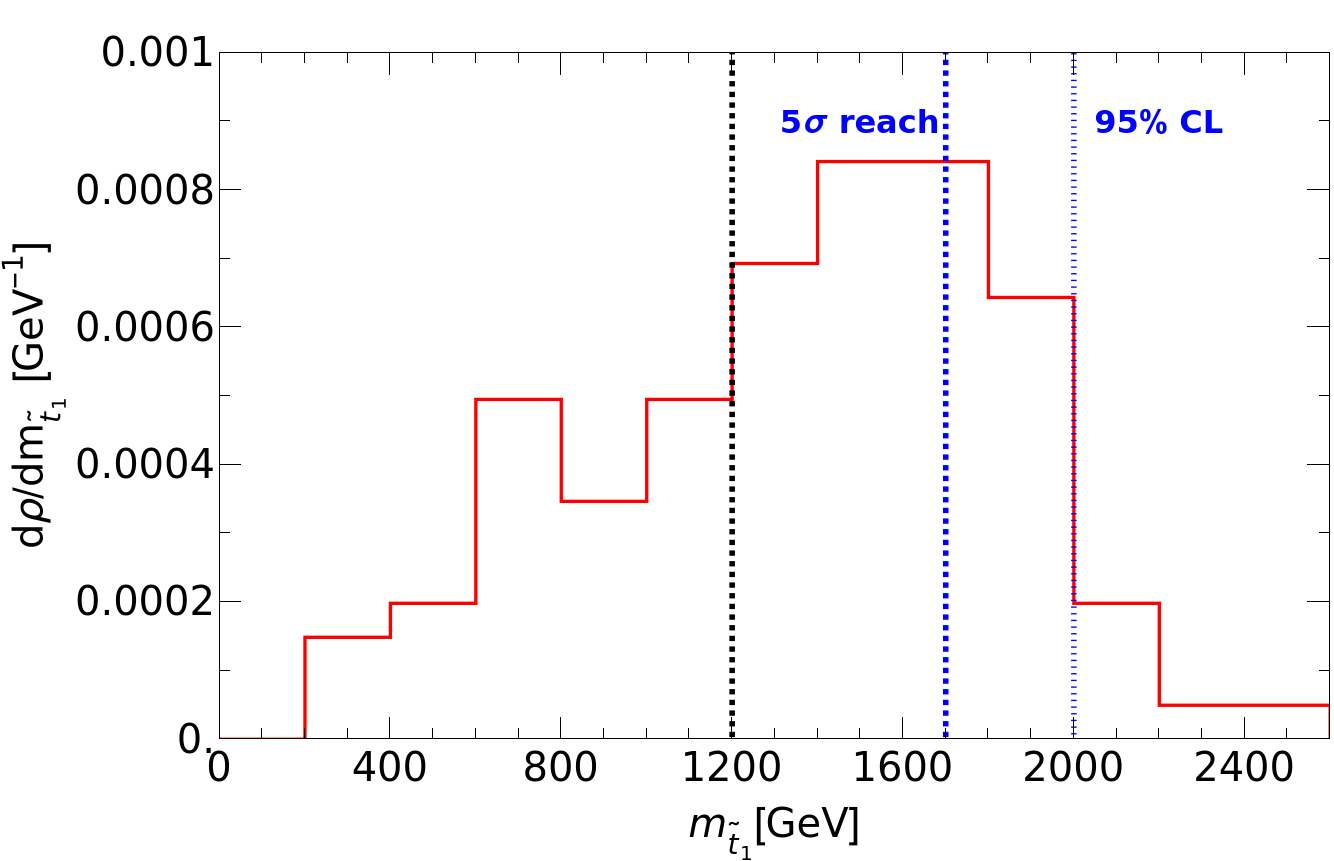}
  \caption{Probability distribution for lighter top squark mass $m_{\tst_1}$
    with an $n=1$ power-law draw to large soft terms.
    We also show the present reach on $m_{\tst_1}$ from LHC Run 2, and
    the expected HL-LHC $5\sigma$ and 95\% CL reach in the figure.
  \label{fig:summary}}
\end{center}
\end{figure}

In spite of the large mixing, the
lighter top-squark is mainly a right-squark, but decays at
$\tst_1\to b\tchi_1^+$ at $\sim 50\%$ and $\tst_1\to t\tchi_{1,2}^0$
at $\sim 25\%$ each.
Thus, we expect top-squark pair production at LHC Run 3 and HL-LHC
to lead to mixed final states
of $b\bar{b}+\eslt$, $t\bar{t}+\eslt$ and $tb+\eslt$.
We evaluated some optimized cuts for each of these channels, and then expect
the top-squark pair production to be revealed as an enhancement in the $m_{T_2}$
distribution at high values of $m_{T_2}$.
We combined the reaches in these three channels to find that HL-LHC
operating at $\sqrt{s}=14$ TeV with 3000 fb$^{-1}$
of integrated luminosity should have a $5\sigma$ discovery reach
to $m_{\tst_1}\sim 1.7$ TeV and a 95\% CL exclusion reach to about
$m_{\tst_1}\sim 2$ TeV.
Now our HL-LHC reach results can be added to Fig. \ref{fig:mhmt1}{\it b})
as a final summary frame: Fig. \ref{fig:summary}.
These HL-LHC reach limits will cover {\it most} (but not all)
of the expected stringy natural parameter space from SUSY on the landscape!

{\it Acknowledgements:} 

This material is based upon work supported by the U.S. Department of Energy, 
Office of Science, Office of High Energy Physics under Award Number DE-SC-0009956 and DE-SC-0017647.


\bibliography{stop}
\bibliographystyle{elsarticle-num}

\end{document}